\documentclass[pdflatex,sn-mathphys-ay]{sn-jnl}

\usepackage{graphicx}%
\usepackage{multirow}%
\usepackage{amsmath,amssymb,amsfonts}%
\usepackage{amsthm}%
\usepackage{mathrsfs}%
\usepackage[title]{appendix}%
\usepackage{xcolor}%
\usepackage{textcomp}%
\usepackage{manyfoot}%
\usepackage{booktabs}%
\usepackage{algorithm}%
\usepackage{algorithmicx}%
\usepackage{algpseudocode}%
\usepackage{listings}%
\usepackage{bm}
\usepackage{subcaption}
\usepackage{bbm}
\usepackage{tabularx}

\usepackage{xr}
\makeatletter
\newcommand*{\addFileDependency}[1]{
\typeout{(#1)}
\@addtofilelist{#1}
\IfFileExists{#1}{}{\typeout{No file #1.}}
}\makeatother
\newcommand*{\myexternaldocument}[1]{%
\externaldocument{#1}%
\addFileDependency{#1.tex}%
\addFileDependency{#1.aux}%
}
\myexternaldocument{supplementary}

\newcommand{\Yt}{\{Y_t\}}
\newcommand{\Xt}{\{\bm{X}_t\}}
\newcommand{\Zt}{\{Z_t\}}
\newcommand{\mut}{\mu(\bm{x}_t)}
\newcommand{\sigt}{\sigma(\bm{x}_t)}

\newcommand{\bw}{\bm{w}}
\newcommand{\bW}{\bm{W}}
\newcommand{\bx}{\bm{x}}
\newcommand{\bX}{\bm{X}}
\newcommand{\bY}{\bm{Y}}
\newcommand{\bZ}{\bm{Z}}

\newcommand{\bz}{\bm{z}}
\newcommand{\bs}{\bm{s}}
\newcommand{\cbr}[1]{\left\{ {#1} \right\}}
\newcommand{\rbr}[1]{\left( {#1} \right)}

\newcommand{\prob}{\operatorname{\mathsf{P}}}

\theoremstyle{thmstyleone}%
\theoremstyle{thmstyletwo}%

\theoremstyle{thmstylethree}%

\definecolor{royalblue(web)}{rgb}{0.25, 0.41, 0.88}

\begin{document}

\title[Article Title]{\centering Analysing Extreme Rainfall via a Geometric Framework\\
\vspace{1mm}
\large EVA (2025) Conference Data Challenge: Team Lancaster Geometric}


\author[1]{\fnm{Ryan} \sur{Campbell}}\email{ryan.campbell@univ-cotedazur.fr}
\equalcont{These authors contributed equally to this work.}

\author[2]{\fnm{Kristina} \sur{Grolmusova}}\email{k.grolmusova1@lancaster.ac.uk}
\equalcont{These authors contributed equally to this work.}

\author[3]{\fnm{Lydia} \sur{Kakampakou}}\email{l.kakampakou1@lancaster.ac.uk}
\equalcont{These authors contributed equally to this work.}

\author*[3]{\fnm{Jeongjin} \sur{Lee}}\email{j.lee58@lancaster.ac.uk}
\equalcont{These authors contributed equally to this work.}


\affil[1]{\orgdiv{Laboratoire J.A. Dieudonn\'{e}}, \orgname{Universit\'{e} C\^{o}te d'Azur}, \orgaddress{\city{Nice}, \country{France}}}

\affil[2]{\orgdiv{STOR-i Centre for Doctoral Training}, \orgname{Lancaster University}, \orgaddress{\city{Lancaster}, \country{United Kingdom}}}

\affil[3]{\orgdiv{School of Mathematical Sciences}, \orgname{Lancaster University}, \orgaddress{\city{Lancaster}, \country{United Kingdom}}}


\abstract{
    Motivated by the EVA 2025 Data Challenge, we address the problem of predicting extreme rainfall in the eastern United States using data from a large ensemble of climate model runs.
    The challenge focuses on three quantities of interest related to the spatial extent and/or temporal duration of extreme rainfall, each requiring extrapolation.
    To tackle these questions, we adopt the recently developed geometric framework for extreme-value analysis, offering substantial flexibility for capturing complex extremal dependence structures and enabling extrapolation across the entire multivariate tail.
    In this work, we focus on the spatial geometric framework for analysing the spatial extent and consider a sampling procedure that retains the temporal information in the data, thereby enabling estimation of the duration of extreme rainfall events.
    We also account for the non-stationary behaviour, arising from topographical and seasonal effects, that commonly characterises extreme weather events in both space and time.
    Using diagnostic metrics, we demonstrate that the proposed model is appropriate for inferring extreme events on this dataset and apply it to estimate target quantities of interest.
}

\keywords{geometric extremes, extremal dependence, extrapolation, non-stationarity, spatial extremes, hydrology}


\maketitle

\section{Introduction}
\label{sec:intro}
Being able to characterise extreme weather events is crucial for the building of defence systems.
A motivating example for our study is Hurricane Helene, which brought catastrophic flooding to the region surrounding Asheville, North Carolina, due in large part to extreme rainfall, resulting in hundreds of fatalities and billions of dollars in damage.
Understanding how often such extreme precipitation occurs and how far it extends spatially is therefore essential for mitigating such losses in the future.

Motivated by the EVA $2025$ Data Challenge, we address questions concerning the spatial extent and/or temporal duration of extreme rainfall, each requiring extrapolation beyond observed levels.
We analyse precipitation output from the CESM2 Large Ensemble Community Project climate model (LENS2, \cite{rodgers2021ubiquity}), taken over a $5\times 5$ spatial grid centred on the location closest to Asheville, NC. 
While these model runs represent plausible climate realisations, they do not replicate actual historical weather events and thus cannot directly reproduce specific storms, motivating the use of statistical models grounded in extreme value theory.

In this work, our main focus is on modelling a spatial process $\cbr{Z(\bs):\bs\in\mathcal{S}\subset\mathbb{R}^2}$ over a gridded spatial domain $\mathcal{S}$ when it attains extreme levels.
To assess the risks associated with extreme behaviour in $Z$, we consider spatial extremes models that enable investigation of the probabilities of events more extreme than those seen to date.
To explore the spatial extent of an extreme event, we begin by considering random variables $Z(\bs)$ and $Z(\bs+h_s)$ at two spatial locations separated by spatial displacement $h_s$.
For marginal distributions $Z(\bs)\sim F$, their joint tail behaviour can be summarised via the so-called tail dependence coefficient
\begin{equation}
\label{eq:s_chi}
    \chi(u;\bs,h_s)=\prob\cbr{F(Z(\bs+h_s))>u\mid F(Z(\bs))>u}, 
\end{equation}
for $u$ close to 1.
If the process is spatially stationary and isotropic, then the coefficient depends only on the spatial separation, i.e., $\chi(u;h_s)$.
When the limit $\lim_{u\rightarrow 1}\chi(u;\bs, h_s)=\chi_{s}$ exists, a value of $\chi_{s}=0$ corresponds to asymptotic independence (AI), indicating extreme events tend to occur on a localised scale \citep{Huser2022}, whereas $\chi_s\in(0,1]$ indicates asymptotic dependence (AD), meaning that the extreme events tend to co-occur across $\mathcal{S}$.
In most environmental application, $\chi(u;\bs, h_s)$ is expected to decay as spatial distance increases.
Figure S.1 in Supplementary Material A.1 displays empirical tail dependence coefficients $\cbr{\hat{\chi}(u;\bs,h_s):u\in[0.95,0.99]}$, plotted against distance.
As expected for rainfall, the spatial extremal dependence decreases over distance.

Environmental extremes often exhibit AI, whereas datasets displaying AD are less common and typically arise in small, geographically homogeneous spatial domains.
Consequently, we require an extreme value framework that can accommodate both classes of extremal dependence, with a particular focus on the AI class.
To this end, we adopt a geometric approach \citep{wadsworth2024statistical,Kakampakou2025} that captures complex joint tail behaviour and allows extrapolation in any direction of the multivariate tail region.
Further, we employ a sampling procedure that preserves the temporal information in the data, thereby enabling estimation of the duration of extreme rainfall events.
Statistical methodology for inference under the geometric framework is relatively recent and has not been tested as extensively as more classical models for multivariate/spatial extremes, especially in higher dimensional settings.
Motivated by this gap in the applied literature, the Data Challenge provides an ideal opportunity to assess its performance in the aforementioned environmental application.

This paper is outlined as follows. Section~\ref{sec:nonstationary} describes the approaches used to account for temporal and spatial non-stationarity in extremes.
Section~\ref{sec:geometric} provides background on the spatial geometric approach.
Section~\ref{sec:model_rainfall} details the implementation of models dealing with non-stationarity, evaluates the performance of the spatial geometric model, and details the estimation of the competition target quantiles via a sampling procedure.
We conclude with a brief discussion in Section~\ref{sec:discussion}.

\section{Modelling non-stationary extremes}
\label{sec:nonstationary}
Environmental data, such as precipitation, frequently exhibit both spatial and temporal non-stationarity in their margins and/or dependence structures.
While it is natural to incorporate external covariates, access to such covariates is not always available and may require prior knowledge.
Extreme climatological phenomena often arise from complex and not fully characterized meteorological processes, making it difficult to identify meaningful covariates.
Acknowledging these practical constraints, we adopt statistical methodologies that can handle non-stationarity with minimal reliance on external covariates. 
Section \ref{sec:temp_nonst} presents methods for addressing temporal non-stationarity in the margins, while Section \ref{sec:spatial_nonst} considers spatial non-stationarity in both marginal and joint extremes.

\subsection{Modelling temporally non-stationary extremes}
\label{sec:temp_nonst}

Let $\Yt$ be a non-stationary process, with an associated sequence of covariates $\Xt$.
Our focus is on modelling the marginal non-stationarity of $\Yt$ through the conditional distribution $Y_t\mid\bm{X}_t=\bm{x}_t$.
We adapt the preprocessing approach proposed by \cite{eastoe2009modelling} for modelling extreme non-stationarity.
The key idea involves two steps: preprocessing the entire data series and then analysing the extremes of the resulting preprocessed data.
The method is expected to effectively remove most of the non-stationarity of $\Yt$, thereby facilitating a subsequent simplified extreme value analysis of the preprocessed series.

We consider a location-scale model for the non-stationary process $\Yt$:
\begin{equation}
\label{eq:location-scale}
    Y_t=\mut+\sigt Z_t,
\end{equation}
where the standardized process $\Zt$ is assumed to be approximately stationary, and $\mut$ and $\sigt$ denote the location and scale parameters, respectively.
We further assume that the bulk of the $\Zt$ distribution is stationary and can be modelled by its empirical distribution.
However, the extremes of $\Zt$ may not conform to those of a stationary series.
We therefore model the extreme values of $\Zt$ using methods for non-stationary extremes.

We define the extremes of $\Zt$ as the exceedances of a high threshold $u$.
The distribution of these excesses, $Z_t-u$, converges to a non-degenerate limiting distribution, the generalized Pareto distribution (GPD).
To account for the non-stationarity of the exceedances, we model the GPD parameters as functions of the covariates \citep{Davison1990}.
The distribution of the excesses, denoted by GPD($\psi_u(\bx),\xi(\bx))$, is given for $z>0$ by
\begin{equation}
\label{eq:gpd}
    P(Z_t>z_t+u\mid z_t>u, \bm{X}_t=\bm{x}_t)=\rbr{1+\frac{\xi(\bx_t)}{\psi_u(\bx_t)}z_t}_+^{-1/\xi(\bx_t)},
\end{equation}
where $a_+=\max(a,0)$, and $\psi_u(\bx_t)$ and $\xi(\bx_t)$ are the scale and shape parameters, respectively.

\subsection{Modelling spatial non-stationary extremes}
\label{sec:spatial_nonst}

We apply the spatial deformation approach of \cite{richards2021spatialdeformation} to address non-stationarity in the spatial extremal dependence.
The method can be implemented prior to fitting a spatial extremes model and does not require prior knowledge of covariates.
The key idea is to use a thin-plate spline --- a smooth mapping function $q(\bs_i)=\bs_i^*$ --- to transform the sampling locations $\bs_1,\ldots,\bs_d$ in the original geographical space (``G-plane") to locations $\bs_1^*,\ldots,\bs_d^*$ in a latent space (``D-plane") in which stationarity is a reasonable assumption.

The parameters of the thin-plate spline are estimated by minimising the Frobenius norm of the discrepancy between the theoretical and empirical pairwise extremal dependence matrices;  $$\min\|\Sigma_X-\widehat{\Sigma}_X\|_F,$$ where $\Sigma_X=[\chi(h_{s}^*)]$ is the theoretical matrix, $h_s^*=\|\bs_i^*-\bs_j^*\|_{i,j=1,\ldots,d}$ denotes the distance between the process at locations $\bs_i^*$ and $\bs_j^*$ in the latent space, and $\widehat{\Sigma}_X=[\hat{\chi}(h_s^*)]$ is its empirical counterpart.
We use a Brown-Resnick process as a parametric model for $\chi(h_s^*)$ due to its flexibility in capturing various extremal dependence structures.

\section{Spatial geometric framework}\label{sec:geometric}

The geometric approach is based on the study of appropriately scaled sampled clouds and their convergence to limit sets \citep{davis1988almost,kinoshita1991convergence,balkema2010meta}.
The shape of the limit set characterises the form of extremal dependence.
Convergence to limit sets requires light-tailed margins; thus, we assume standard exponential margins throughout, allowing us to focus on extremal dependence in the positive orthant. 
Such marginal transformation is standard in extreme value dependence modelling.

In a spatial context, consider a stationary spatial process $\{Z(\bs):\bs\in\mathcal{S}\}$ with standard exponential margins.
Let $\bZ:=(Z(\bs_1),\ldots,Z(\bs_d))^\top$ be the $d$-dimensional representation of ${Z(\bs)}$ at the observed spatial locations, $\bs_1,\ldots,\bs_d$. 
We let $\bZ_t$, $t=1,\ldots,n$, be independent copies of the random vector $\bZ:=(Z(\bs_1),\ldots,Z(\bs_d))^\top$.
The scaled $n$-point sample cloud $N_n=\cbr{\bZ_1/\log n,\ldots,\bZ_n/\log n}$
converges onto the limit set $G$ under conditions given by \cite{balkema-nolde-2010}.
The limit set is characterised by a gauge function $g_{\bZ}(\bz)$ that is homogeneous of order $1$, and can be written as $G=\cbr{\bz\in\mathbb{R}_+^d:g_{\bZ}(\bz)\le 1}$.
To facilitate determination of the limit set for a given distribution, \cite{nolde2014geometric} and \cite{nolde2022linking} provide a sufficient condition for convergence in terms of the Lebesgue density of $\bZ$, denoted by $f_{\bZ}$,
\begin{equation*}
    \lim_{l\rightarrow \infty}\frac{-\log f_{\bZ}(l\bz)}{l}=g_{\bZ}(\bz),\quad \bz\in[0,\infty]^d.
\end{equation*}

Assuming that the random vector $\bZ$ admits a density with exponential margins, we adopt the statistical framework of \cite{wadsworth2024statistical}.
We apply the radial-angular transformation $\bZ\mapsto (R,\bW)$ with 
\begin{equation*}
    R=\sum_{\bs_i \in\mathcal{S}}Z(\bs_i),\quad \bW=\bZ/R,
\end{equation*}
where $\bW$ lies on $\Theta_+^{d-1}=\cbr{\bw\in\mathbb{R}_+^d:\|\bw\|=1}$.
We consider $\bZ$ to be {\em extreme} if $R>r_{\tau}(\bw)\,|\,\bW=\bw$, where $r_\tau(\bw)$ is a high threshold.
We further assume that the conditional distribution of $R\mid \bW$ above a high threshold $r_\tau(\bw)$ follows a truncated gamma distribution, denoted TG,
\begin{equation}
\label{eq:TG}
    R\mid [\bW=\bw, R>r_{\tau}(\bw)]\sim \text{TG}\rbr{\lambda d, g_{\bZ}(\bw)},
\end{equation}
where $\lambda d$ with $\lambda>0$, and $g_{\bZ}(\bw)$ are the gamma shape and rate parameters, respectively. 
Although the theoretical shape parameter in the truncated gamma model~\eqref{eq:TG} corresponds to $d$ for many parametric gauge functions, we allow for additional flexibility by estimating it through the parametrisation $\lambda d$ with $\lambda>0$, which also improves numerical stability \citep{Kakampakou2025}.
In this work, we parametrise the gauge function as $g_{\bZ}(\bw)=g_{\bZ}(\bw;\bm{\theta})$ and estimate its parameters using the gamma approximation.
For further statistical justification, we refer to \cite{wadsworth2024statistical}.

To extend this framework to the spatial setting, we adopt the spatial geometric approach of \cite{Kakampakou2025}.
In particular, we consider the generalised Gaussian gauge function
\begin{equation}
\label{eq:glg}
    g_{\bZ}(\bw;\gamma) 
    =
    \left[(\bw^{1/{\gamma}})^\top \Sigma^{-1}\bw^{1/{\gamma}}\right]^{{\gamma}/{2}},\quad \gamma\in(0,\infty),
\end{equation}
where $\Sigma$ denotes a correlation matrix.
This specification generalises the standard Gaussian form induced by a spatial Gaussian process by adding an additional parameter, and its flexibility in capturing the extremal dependence structure is demonstrated in \cite{Kakampakou2025}.
Specifically, we consider an isotropic and stationary correlation matrix $\Sigma(\phi,\kappa)$ with powered exponential components in space
\begin{equation}
    [\Sigma(\phi,\kappa)]=\exp\cbr{\rbr{-h_s/\phi}^\kappa},\quad \phi\times\kappa\in(0,\infty)\times(0,2],
\end{equation}
where $\phi$, $\kappa$ are the range and shape parameters, respectively, and $h_s=\|\bs_i-\bs_j\|_2$ denotes the Euclidean distance between locations $\bs_i,\bs_j\in \mathcal{S}=\cbr{(j,k)}$ for $j,k=1,\ldots,d$.
The parameter $\phi$ controls the strength of spatial dependence with distance, while $\kappa$ controls the properties of the
spatial process and, more specifically, the
roughness of its realizations \citep{davison2012statistical}.

\section{Extreme rainfall in the Eastern U.S.}
\label{sec:model_rainfall}
\subsection{Exploratory data analysis}

\subsubsection{Temporal non-stationarity}
\label{sec:detrending}

Inference for the temporally non-stationary extremes model involves a two-step procedure: first estimating the location-scale parameters in~\eqref{eq:location-scale}, and then modelling the tail of the standardized series in~\eqref{eq:gpd} by means of a non-stationary GPD.
Following \cite{eastoe2009modelling}, we assume a Gaussian family for modelling the location, $\mu(\bx_t)$, and scale, $\sigma(\bx_t)$, in the underlying bulk distribution of $\Yt$, as it provides robustness to extreme observations in the tails.

We use a generalized additive model (GAM) to flexibly represent the parameters $\mu(\bx_t)$ and $\log(\sigma(\bx_t))$ through both linear and cyclic trends as in \citet{Murphy-Barltrop2024}.
To account for residual temporal dependence arising from missing covariates, particularly those linked to climate processes, we also include auto-regressive (AR) terms in the mean.
Specifically, we allow the model parameters to depend linearly on precipitation from the previous three days, along with covariates describing the daily cycle, long-term trends in both the mean and variability.
All covariates were found to be significant ($p$-value $<0.01$), and including up to three lagged precipitation terms yielded the best model fit in terms of the lowest Akaike information criterion (AIC).
For inference, we use restricted maximum likelihood estimation to estimate the mean $\mu(\bx_t)$ and scale parameters $\sigma(\bx_t)$.

Figure S.2 in Supplementary Material A.1 displays the estimated location and scale parameters for the first variable (corresponding to location $\bm{s}_1$) in climate model run 1, shown as a representative example.
The estimated location parameter indicates that the daily precipitation remains relatively flat, with slightly lower levels in summer and just before the winter months.
The estimated scale parameter exhibits a clear seasonal pattern, characterised by lower variability in summer and higher variability in winter.
Overall, the long-term variability fluctuates over time, while the average level remains relatively constant.

Next, we model the tail of the standardized series, which is approximately stationary.
Standard diagnostic plots such as mean residual life (omitted) suggest that the 80\% quantile gives an adequate threshold $u(\bs_j)$ for $j=1,\ldots,25$, consistent with the truncated gamma threshold level chosen in~\eqref{eq:threshold_tau}.
Given random samples $\bZ_t=(Z_t(\bs_1),\ldots,Z_t(\bs_d))^\top$, $t=1,\ldots,n=60225$, we standardise the margins to exponential form using a semi-parametric marginal transformation applied at each site $\bs_j$, $j=1,\ldots,25$, followed by the inverse probability integral transform.
Specifically, for each preprocessed variable $Z(\bs_j)$, we estimate the marginal distribution using an empirical distribution via rank transformation below the threshold $u(\bs_j)$, and a non-stationary GPD above the threshold \citep{TawnColesSemipar}.
Let $\widehat{F}_{j}(Z_t(\bs_j))$ denote the estimator of the marginal distribution function $F_j(z)=\prob(Z_t(\bs_j)\le z).$
The resulting fitted marginal distribution is
\begin{align*}
    \widehat{F}_j(z_t(\bs_j))=
    \begin{cases}
        1-\rbr{1+\frac{\hat{\xi}(\bx_t)}{\hat{\psi}_u(\bx_t)}z_t(\bs_j)}_+^{-1/\hat{\xi}(\bx_t)},\quad &z_t(\bs_j) > u(\bs_j)\\
        \text{R}_{t,j} / (n+1),\quad &z_t(\bs_j) \le u(\bs_j),
    \end{cases}
\end{align*}
where $\text{R}_{t,j}=\sum_{\ell=1}^n \mathbb{I}(Z_\ell(\bs_j)\le Z_t(\bs_j))$ is the rank of $Z_t(\bs_j)$ among $Z_1(\bs_j),\ldots,Z_n(\bs_j)$ at site $\bs_j.$
After estimating the margin, we assume the margins are stationary over time and transform them to the standard exponential scale via $-\log(1-\widehat{F}_j(Z_t(\bs_j))$, $j=1,\ldots,25.$

\subsubsection{Spatial non-stationarity}
\label{sec:spatial_nonst_application}

\begin{figure}[t!]
    \centering
    \includegraphics[trim={0 1cm 0 2cm},clip,width=0.44\linewidth]{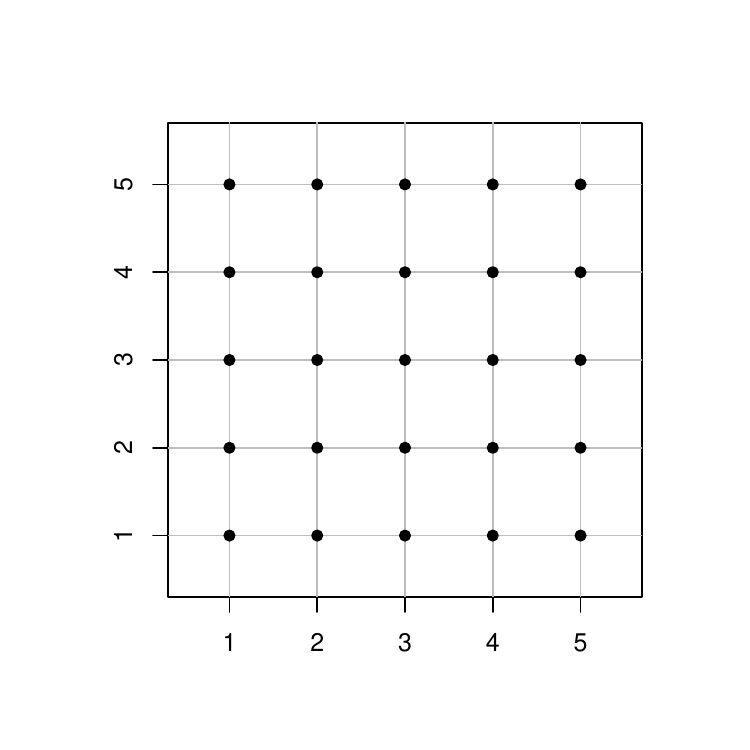}
    \includegraphics[trim={0 1cm 0 2cm},clip,width=0.44\linewidth]{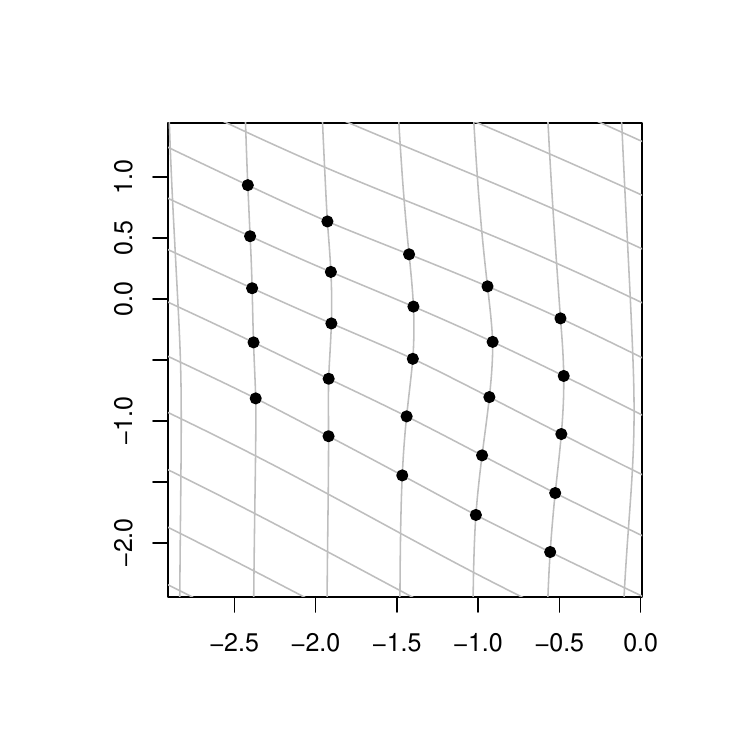}
    \caption{Left: Original grid coordinates (G-plane) of the eastern U.S. rainfall data. Right: Deformed grid coordinates (D-plane) obtained using data from run 1 of the climate model.}
    \label{fig:stat-coords}
\end{figure}

Exploratory analysis shows clear anisotropy and non-stationarity across the G-plane.
This is linked to geographical features, and indicates that rainfall extremes vary across our spatial domain: the probability of extreme rainfall at one location may differ from that at another.
To account for this, we deform the original grid $\mathcal{S}$ from the climate model output into a new set of locations in the D-plane where stationarity is assumed to hold.
The deformation is applied to the {\em preprocessed} data, after accounting for temporal non-stationarity as described in Section~\ref{sec:detrending}.
Using the \texttt{R} package \texttt{sdfExtreme}, and a chosen set of {\em anchor} points to estimate the deformation, Figure~\ref{fig:stat-coords} displays the resulting G-plane and D-plane for climate model run 1.
The plot exhibits clear differences in spatial behaviour between the upper-left and lower-right portions of the grid.

As outlined in Section~\ref{sec:spatial_nonst}, we estimate the mapping parameters by minimising the distance between the theoretical and empirical tail dependence coefficients.
Under AD, we assume that $\chi(u,h_s)=\chi(h_s)$ for $u$ close to 1.
Figure S.4 in Supplementary Material A.2 shows empirical estimates $\hat{\chi}(u;h_s)$ at $u=0.99$, plotted against distance before and after deformation.
The reduced variability in the empirical $\hat{\chi}(u;h_s)$ values after deformation indicates improved stationarity in the D-plane, with both anisotropy and non-stationarity reduced.

Following the suggestions of \cite{richards2021spatialdeformation}, 
we set the number of {\em anchor} points to a quarter of the 25 sampling locations to avoid overfitting in the deformation step.
Different choices of anchor points lead to slightly different D-planes, though using a regular three-point spacing space tends to produce broadly similar results.
Note that there is no formally optimal strategy for choosing anchor site in this setting.

\subsection{Fitted geometric spatial model}
\label{sec:fittedmodel}
We follow the statistical inference and prediction procedure described in \citet{wadsworth2024statistical} and \citet{Kakampakou2025}, including a brief summary for completeness.

\subsubsection{Statistical inference}
\label{sec:inference}
To fit the truncated gamma model in~\eqref{eq:TG} above a high threshold, we first need to calculate a high threshold $r_{\tau}(\bw)$ of the conditional distribution $R\mid \bW=\bw$ such that $\bar{F}(r_{\tau}(\bw)\mid \bw)=1-\tau$, where $\bar{F}(r\mid\bw)$ denotes the gamma survival function of $R\mid\bW=\bw$.
We can estimate this threshold $r_\tau(\bw)$ empirically or via additive quantile regression, but both approaches are limited to lower dimensions $(d\approx 5)$ due to the specific support of $\bW$ on $\Theta_{+}^{d-1}$.
In this work, we approximate $r_{\tau}(\bw)$ for $\tau$ close to 1 using the following approximation
\begin{equation*}
    r_\tau(\bw)\approx C_\tau/g(\bw;\bm{\theta}),
\end{equation*}
where $C_\tau$ solves $\int_0^{C_\tau}\Gamma(\lambda)^{-1}v^{\lambda-1}\exp(-v)dv=\tau$ for $\tau$ close to 1.
Thus, the properly scaled threshold can be interpreted as a non/semi-parametric estimate of the gauge function \citep{wadsworth2024statistical}.

In a spatial context, \cite{Kakampakou2025} adopt this approximation and estimate $C_\tau$ via a pairwise composite likelihood approach.
Following their approach, we begin by estimating the parameters $(\phi,\kappa)$ of a bivariate standard Gaussian gauge function, fixing $\gamma=2$ in~\eqref{eq:glg}, using all $d\choose 2$ pairs.
Based on graphical diagnostic assessments of tail fit, we set $\tau=0.8$; in unreported investigations, this choice provided satisfactory performance relative to the other quantiles considered.
We then obtain an estimate $\widehat{C}_\tau$ such that the proportion of threshold exceedances is as close to $1-\tau$ as possible.
Substituting the parameter estimates into the standard Gaussian gauge function, the threshold is set as
\begin{equation}
\label{eq:threshold_tau}
r_{\tau}(\bw)=\widehat{C}_\tau / g(\bw;\hat{\phi},\hat{\kappa}).    
\end{equation}
For more details on this approach, see \citet{Kakampakou2025}. 

Using threshold exceedances $(r_i,\bw_i)$, $i=1,\ldots,\tilde{n}$ such that $r_i>r_\tau(\bw)$, we fit the truncated gamma model with the $d$-dimensional {\em generalised} Gaussian gauge function via maximum likelihood.
The likelihood function is
\begin{equation}
\label{eq:likelihood}    L(\lambda,\bm{\theta};r,\bw)=\prod_{i=1}^{\tilde{n}}\frac{g_{\bZ}(\bw_i;\bm{\theta})^{\lambda d}}{\Gamma(\lambda d)}\frac{r_i^{\lambda d-1}\exp\rbr{-r_ig_{\bZ}(\bw_i;\bm{\theta})}}{\bar{F}(r_\tau(\bw_i);\lambda d, g_{\bZ}(\bw_i,\bm{\theta}))},
\end{equation}
where $\bar{F}(\cdot;\lambda d, g_{\bZ}(\bw;\bm{\theta}))$ denotes the gamma survival function.
The maximum likelihood estimates for each of the four climate model runs are presented in Table~\ref{tab:MLEs}.
\begin{table}[h!]
    \centering
    \begin{tabularx}{0.77\textwidth}{>{\centering\arraybackslash}X|>{\centering\arraybackslash}X|>{\centering\arraybackslash}X|>{\centering\arraybackslash}X|>{\centering\arraybackslash}X}
         Run &  $\hat{\lambda}$ &  $\hat{\phi}$ &  $\hat{\kappa}$ &  $\hat{\gamma}$ \\ \hline
        1 &0.224 &0.830 &1.89 &1.16 \\
        2 &0.224 &0.828 &1.89 &1.17 \\
        3 &0.219 &0.810 &1.92 &1.11 \\
        4 &0.218 &0.811 &1.91 &1.13\\
    \end{tabularx}
    \caption{Maximum likelihood estimates (in 3 s.f.) of the truncated gamma model in~\eqref{eq:TG} for each of the four climate model runs.}
    \label{tab:MLEs}
\end{table}

\subsubsection{Extrapolation}
\label{sec:extrapolation}

A central benefit of the geometric approach is that the fitted model allows us to extrapolate in any direction of the multivariate tail and thereby estimate tail probabilities beyond the observed data.
To estimate tail probabilities, we require samples of $\bZ$ that fall in an extreme set.
Letting $R'=R/r_\tau(\bW)$, we estimate the probability of such extreme sets using the equation
\begin{equation}
\label{eq:tailprob}
    \prob(\bZ\in B)=\prob(\bZ\in B\mid R'>1)\prob(R'>1),
\end{equation}
for any set $B$ lying in the extreme region $\{\bz\in\mathbb{R}_+^d:\sum_{j=1}^d z_j>r_\tau(\bz/\sum_{j=1}^d z_j)\}$.
We estimate the first probability in~\eqref{eq:tailprob} empirically using samples drawn from the distribution of $\bZ\mid R'>1$.
Each simulated point is obtained by combining two components:
\begin{itemize}
    \item Draw an angular component $\bw^*$ from the empirical distribution of $\bW \mid R'>1$;
    \item Draw a radial component $r^*$ from the fitted truncated gamma model of $R\mid [\bW=\bw, R>r_\tau(\bw^*)]$ conditional upon $\bw^*.$
\end{itemize}
The resulting simulated point is $\bz^*=r^*\bw^*$.
The second probability in~\eqref{eq:tailprob} is obtained empirically as the proportion of observations satisfying $R'>1$, which should be approximately $1-\tau$.

To extrapolate beyond the observed level, we extend the representation in~\eqref{eq:tailprob} and draw samples from $\bZ\mid R'>k$ with a suitably chosen $k>1$.
The corresponding decomposition is
\begin{equation}
\label{eq:extrapolation}
\prob(\bZ\in B)=\prob(\bZ\in B\mid R'>k)\prob(R'>k \mid R'>1)\prob(R'>1),
\end{equation}
where $k$ is chosen such that $B\subset \{\bz\in\mathbb{R}_+^d:\sum_{j=1}^d z_j > kr_\tau(\bz/\sum_{j=1}^d z_j)\}$.
As before, draws of the radial component are obtained directly from the fitted truncated gamma $R\mid [\bW=\bw, R>kr_\tau(\bw)]$.
Sampling from $\bW\mid R'>k$ is challenging for large $k$, but we can obtain approximate samples by re-weighting an empirical sample from $\bW\mid R'>1$ via importance weights
\begin{equation*}
    IW(\bw):=\frac{\bar{F}(kr_\tau(\bw);\hat{\lambda} d,g_{\bZ}(\bw;\hat{\bm{\theta}}))}{\bar{F}(r_\tau(\bw);\hat{\lambda} d,g_{\bZ}(\bw;\hat{\bm{\theta}}))}
\end{equation*}
The middle probability $\prob(R'>k\mid R'>1)$ is then estimated by $\frac{1}{\tilde{n}}\sum_{i=1}^{\tilde{n}}IW(\bw_i)$.
For details, we refer to \cite{wadsworth2024statistical}.

\subsubsection{Diagnostics}

\begin{figure}[th!]
    \centering
    \includegraphics[trim={0.5cm 0.5cm 1cm 1cm},clip,width=0.24\linewidth]{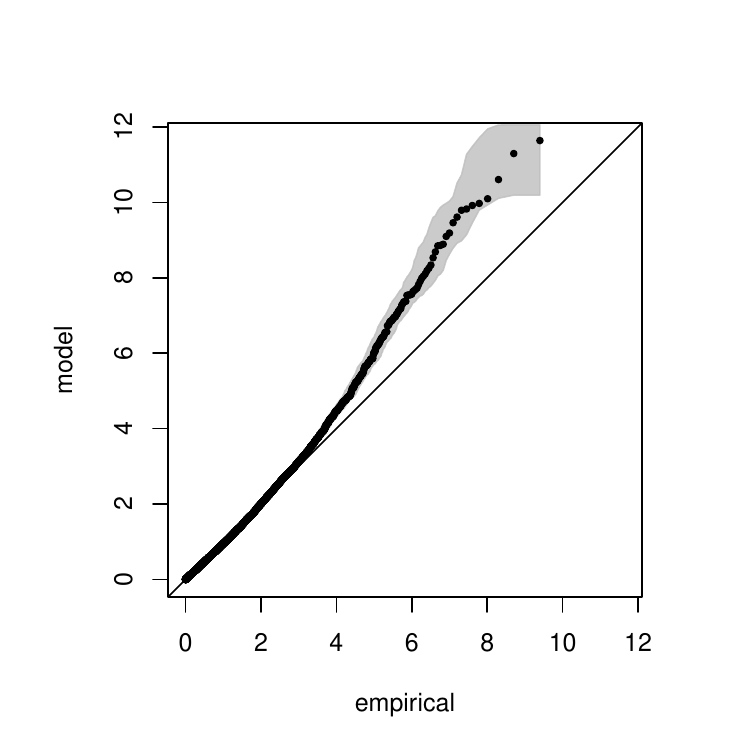}
    \includegraphics[trim={0.5cm 0.5cm 1cm 1cm},clip,width=0.24\linewidth]{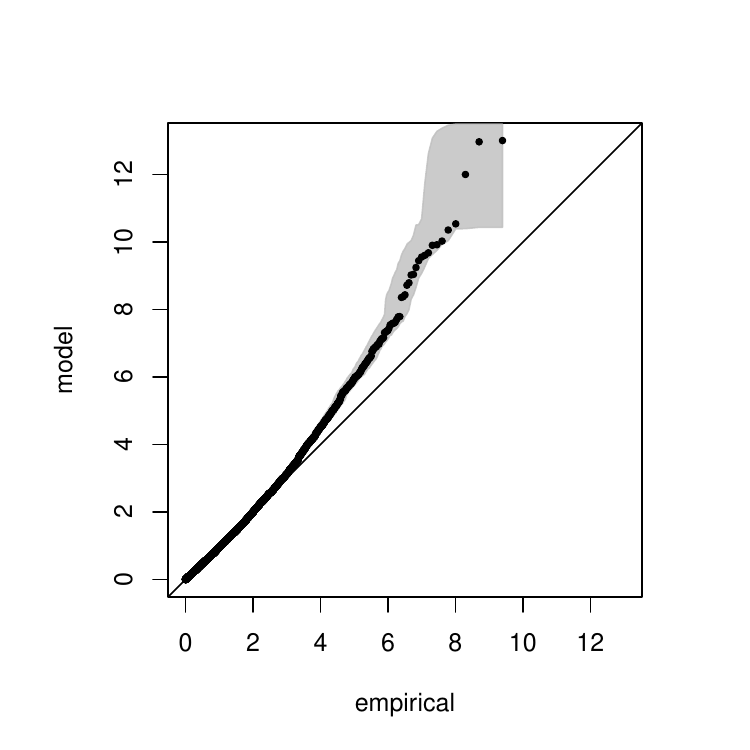}
    \includegraphics[trim={0.5cm 0.5cm 1cm 1cm},clip,width=0.24\linewidth]{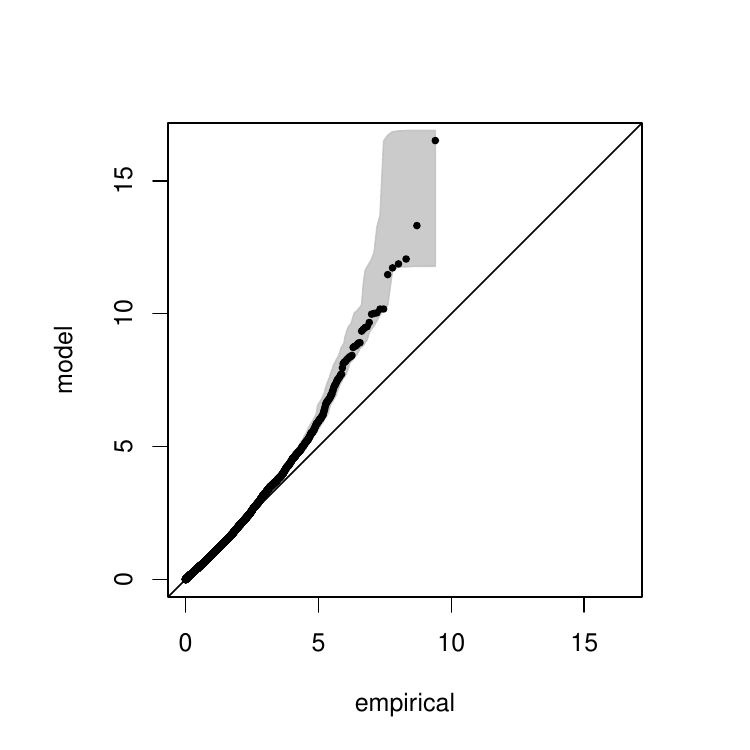}
    \includegraphics[trim={0.5cm 0.5cm 1cm 1cm},clip,width=0.24\linewidth]{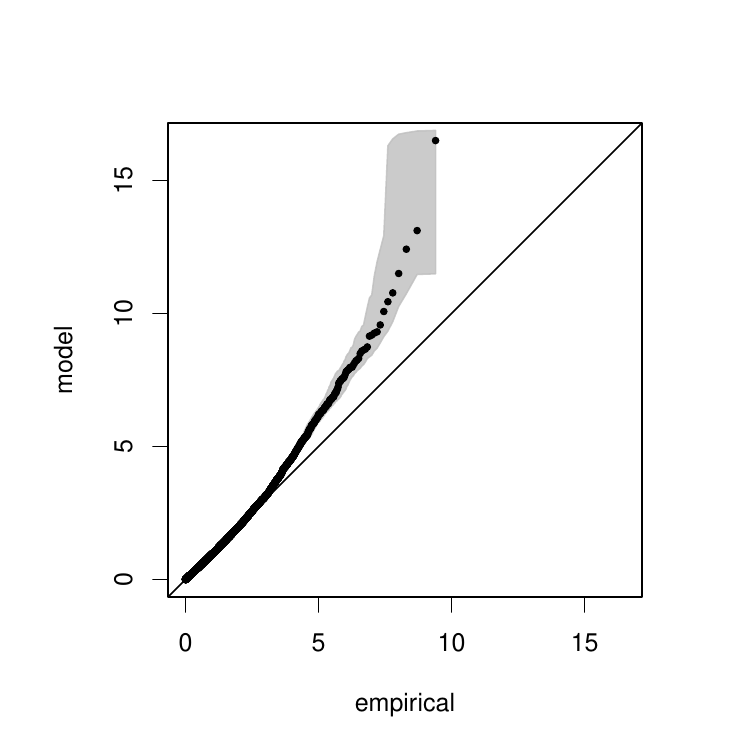}
    \caption{QQ plots for the truncated gamma model fit using data from runs 1 to 4 (left to right) of the climate model. The grey regions indicate 95\% bootstrap confidence intervals.}
    \label{fig:QQ}
\end{figure}

We assess the goodness of fit from the likelihood-based inference using Probability-Probability (PP)  and Quantile-Quantile (QQ) plots \citep{wadsworth2024statistical}.
The fitted truncated gamma distribution above the high threshold $r_\tau(\bw)$ is 
\[
\Tilde{F}(r\mid \bw,r_\tau(\bw)):=1-\bar{F}(r;\hat{\lambda}d,g_{\bZ}(\bw,\hat{\bm{\theta}})\bar{F}(r_{\tau}(\bw);\hat{\lambda}d,g_{\bZ}(\bw;\hat{\bm{\theta}}))^{-1},
\]
with $\bar{F}$ defined in~\eqref{eq:likelihood}. 
For threshold exceedances $(r_i,\bw_i)\mid\{r_i>r_\tau(\bw_i)\}$, $i=1,\ldots,\tilde{n}$, the PP plot is formed from the set of points 
\begin{equation*}
 \cbr{(i/(\tilde{n}+1),u_{(\tilde{n}-i+1)}},\quad u_i=\Tilde{F}(r_i\mid \bw_i,r_\tau(\bw_i)),
\end{equation*}
where $u_{(1)}\ge \cdots \ge u_{(\tilde{n})}$ denote the ordered values.
For each of the four climate model runs, PP plots with 95\% bootstrap confidence intervals are given in Figure S.5 Supplementary Material A.3.
The points appear to lie along the diagonal, indicating an overall satisfactory fit of the truncated gamma model.
To evaluate the tail behaviour of the fitted gamma model, we construct QQ plots on the exponential scale via the points 
\[\cbr{-\log(1-i/(\tilde{n}+1)),-\log(1-u_i)},\]
shown in Figure~\ref{fig:QQ} with 95\% bootstrap confidence intervals.
In it, we see that while we generally have some agreement between the empirical and model-based quantiles, there is a deviation from the diagonal as we examine further into the tails. This is to be expected, as we did not directly model temporal non-stationarity in our truncated gamma model, something we wish to explore in future work.

\begin{figure}[th!]
    \centering
    \includegraphics[trim={0.5cm 0.5cm 1cm 1cm},clip,width=0.24\linewidth]{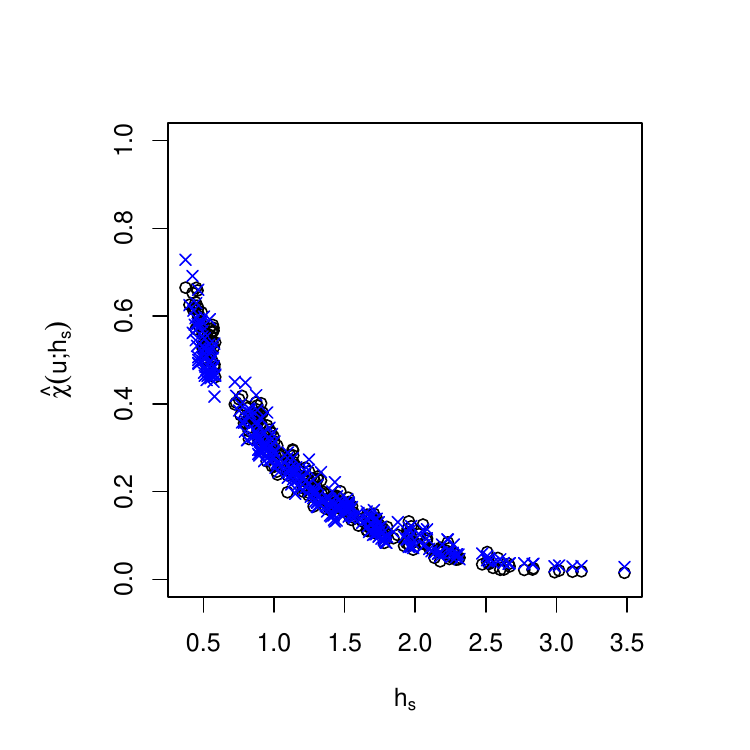}
    \includegraphics[trim={0.5cm 0.5cm 1cm 1cm},clip,width=0.24\linewidth]{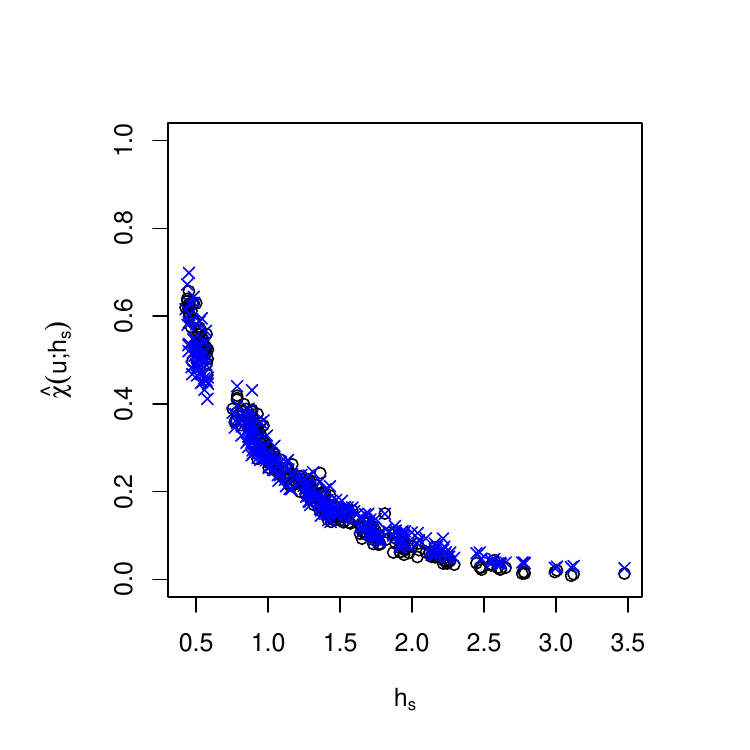}
    \includegraphics[trim={0.5cm 0.5cm 1cm 1cm},clip,width=0.24\linewidth]{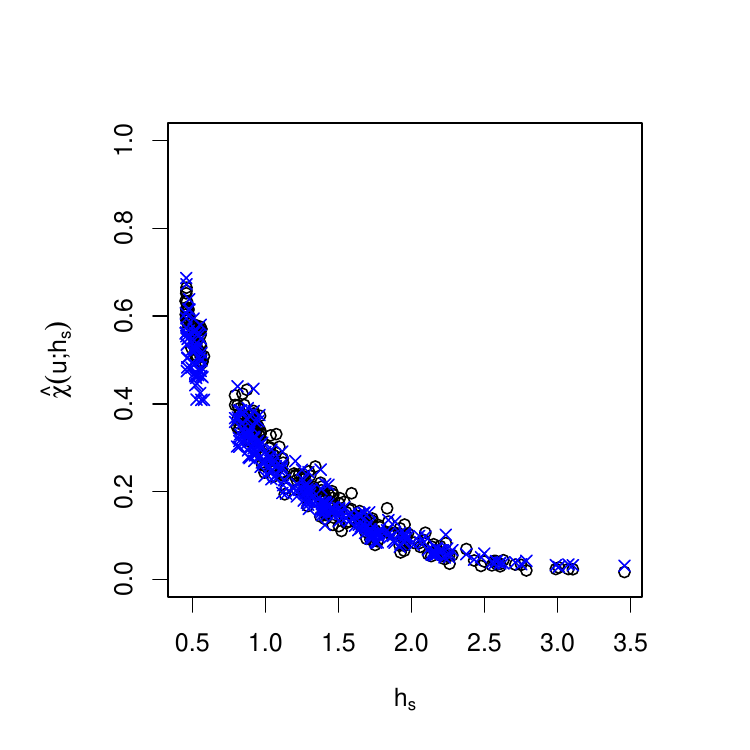}
    \includegraphics[trim={0.5cm 0.5cm 1cm 1cm},clip,width=0.24\linewidth]{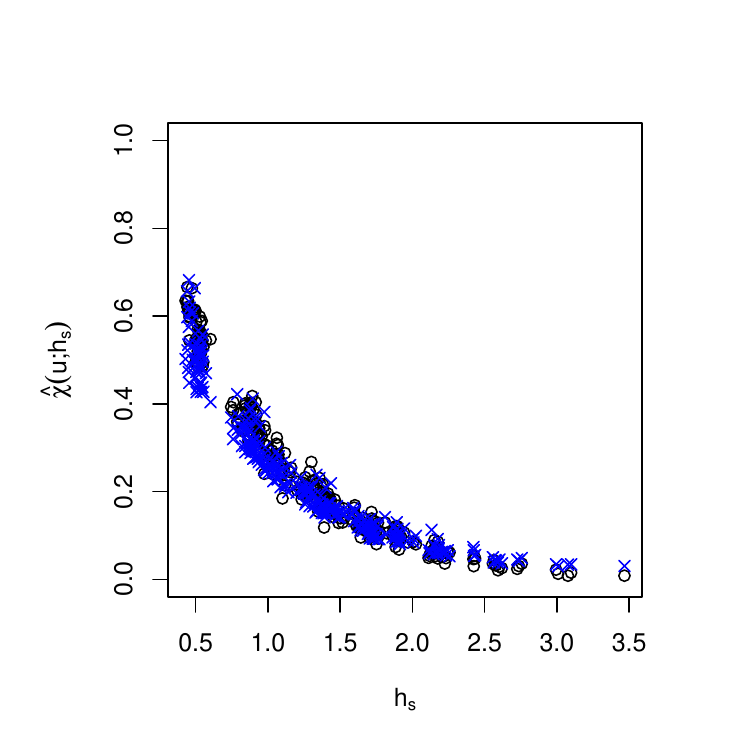}
    \caption{Estimates of $\chi(0.99;h_s)$ for models fitted on the deformed grid using data from runs 1 to 4 (left to right) of the climate model. Empirical estimates are denoted by the black circles ($\circ$), and model-based estimates are given by the blue crosses (\textcolor{blue}{$\times$})}. 
    \label{fig:run1-pwchi}
\end{figure}

Additional to PP and QQ plots, we assess the tail behaviour of the fitted model by comparing empirical and model-based pairwise tail dependence coefficients in the D-plane, where $\chi(u,h_s)$ depends only on spatial separation $h_s=\|\bs_i-\bs_j\|_2$.
For $Z_j\sim F$ with exponential margins, the spatial tail dependence coefficient in~\eqref{eq:s_chi} can be written as
\begin{equation}
\label{eq:chi_general}
\chi(u;\bs_i,\bs_j)=\prob\rbr{F(Z(\bs_i))>u,F(Z(\bs_j))>u}/(1-u),\quad u\in(0,1),
\end{equation}
for $i,j\in\cbr{1,\ldots,d}$. 
Model-based estimates are obtained by simulation from the fitted gamma model with a suitably chosen set $B$.
The numerator in \eqref{eq:chi_general} can be expressed as $\prob(\bm{Z}\in B)$, where the extreme region is given by
\begin{equation*}
B=(0,\infty)^{i-1}\times (-\log(1-u),\infty)^i\times(0,\infty)^{j-i-1}\times(-\log(1-u),\infty)^j\times(0,\infty)^{d-j}.
\end{equation*}
This defines a region above $-\log(1-u)$ in components $i$ and $j$ of the $d$-dimensional positive orthant.
We use the extrapolation procedure described in Section~\ref{sec:extrapolation} with fixed extrapolation level $k=1$, to estimate $\prob(\bm{Z}\in B)$, and therefore $\chi(u,h_s)$.
Figure~\ref{fig:run1-pwchi} compares these model-based tail dependence estimates with empirical tail dependence coefficients at $u=0.99$. Overall, we see good agreement with the empirical estimates across the climate model runs, indicating our model's potential to characterise the joint tails of the climate model data.

\subsection{Estimating the frequency of extreme rainfall events}\label{sec:CTQ-est}

Using the generated exceedances $(r_i,\bw_i)\mid\{r_i>r_\tau(\bw_i)\}$, $i=1,\ldots,\tilde{n}$ obtained through the extrapolation procedure described in Section~\ref{sec:extrapolation}, we estimate the following competition target quantities (CTQs):
\begin{enumerate}
    \item CTQ1: Expected number of times all 25 locations exceed 1.7 Leadbetters.
    \item CTQ2: Expected number of at least 6 of the 25 locations exceed 5.7 Leadbetters.
    \item CTQ3: Expected number of at least 3 of the 25 locations exceed 5 Leadbetters for a run of at least two consecutive days.
\end{enumerate}
Given that we consider observations that have been transformed to standard exponential margins, we begin by converting the values of interest from the Leadbetter scale to the exponential scale for each variable in each climate model run.
As the marginal distribution depends on covariates $\bX_t=\bx_t$ through the GAM and GPD, the quantile function is not readily obtainable in closed form.
We therefore approximate the corresponding exponential-scale value as follows: we identify the covariate value associated with the observation closest to the original threshold, retrieve the corresponding preprocessed data point, and then locate the nearest value in the tail of the preprocessed observation and transform it to the exponential scale via $-\log(1-u^*)$ where $u^*$ denotes the fitted GPD evaluated at the selected point.

Let $\bm{q}_v:=(q_v(\bs_1),\ldots,q_v(\bs_{25}))^\top$ be the approximate threshold vector on the exponential scale for the $v$-th target quantity for $v=1,2,3$.
Let $[d]:=\{1,\ldots,25\}$ be the index set.
We define the three target probabilities to estimate and the associated extreme sets as follows:
\begin{enumerate}
    \item The probability that all 25 locations exceed a high threshold is
    \begin{equation*}
        \prob(\bZ\in B_1)=\prob\rbr{\bigcap_{j\in[d]}\cbr{Z(\bs_j)>q_1(\bs_j)}}
    \end{equation*}
    The extreme set is defined as
    $$B_1=\prod_{j\in[d]}(q_1(\bs_j),\infty).$$
    \item The probability that at least 6 of the 25 sites exceed a high threshold is derived via the inclusion-exclusion principle as
    \begin{equation*}
        \prob(\bZ\in B_2)=\sum_{r=6}^{25} (-1)^{r-6} \sum_{J\subset[d],|J|=r}{r-1\choose 5}\prob(\bZ_J>\bm{q}_{2,J}),
    \end{equation*}
    where $\bZ_J$ denotes the sub-vector of $\bZ$ indexed by $J$.
    The extreme set is defined as the union of all subspaces where any subset of size 6 exceeds the threshold.
    Let $\mathcal{C}_6=\{C\subset[d]:|C|=6\}$.
    The extreme set is
    \begin{equation*}
     B_2=\bigcup_{C\in\mathcal{C}_6}\cbr{\prod_{j\in C}(q_2(\bs_j),\infty)\times\prod_{j\in [d]\setminus{C}}(0,\infty)}.   
    \end{equation*}
    Note that we only need to take the union over subsets of cardinality 6. 
    Any realization with 7 or more exceedances is contained in this union.
    \item 
    Let $\bY(\bs_j):=\min(Z_t(\bs_j),Z_{t+1}(\bs_j)).$
    The probability that at least 3 sites exceed 5 for a run of at least two days is
    \begin{equation*}
        \prob(\bY\in B_3)=\sum_{r=3}^{25}(-1)^{r-3}{r-1 \choose 2}\sum_{J\subset [d],|J|=r}\prob\rbr{\bY_J>\bm{q}_{3,J}}.
    \end{equation*}
    The extreme set is
    \begin{equation*}
    B_3=\bigcup_{C\in\mathcal{C}_3}\cbr{\prod_{j\in C}(q_3(\bs_j),\infty)\times\prod_{j\in [d]\setminus{C}}(0,\infty)}.    
    \end{equation*}
\end{enumerate}

To estimate the target probabilities, we consider the extrapolation equation~\eqref{eq:extrapolation}.
For each extreme set $B_i$, $i=1,2,3,$ we want the largest $k$ such that $B_i\subset \{\bz\in\mathbb{R}_+^d:\sum_{j=1}^d z_j > kr_\tau(\bz/\sum_{j=1}^d z_j)\}$, ensuring that a sufficient number of simulated points fall in $B_i$.
The condition depends on the geometry of $B_i$ and the form of $r_\tau(\bz/\sum_{j=1}^d z_j)$.
For $B_1$, the extreme region of interest has a relatively simple upper-right hypercube shape. 
In contrast, the regions associated with $B_2$ and $B_3$ are more complex and irregular.
Thus, we adopt a practical strategy for selecting $k$ in these cases.
We consider the set of non-exceedances $\{\bm{z}_i: \sum_{j=1}^{d} z_{i,j} < r_{\tau}(\bm{z}_i/\sum_{j=1}^{d} z_{i,j})\}$, scale each by $k$, and search over candidate values $k \in [1, 4]$.
We then select the largest $k$ for which none of the scaled non-exceedances fall in $B$.
We found that this heuristic method is effective for complex extreme sets, provided that the number of simulated points is sufficiently large.

Once an appropriate $k$ is selected for each target quantity, we simulate $10^6$ extremal points $\bz_i^*$, using the procedure described in Section~\ref{sec:extrapolation}.
In particular, for the third target probability, we need to account for temporal dependence in the series.
To do so, we sample temporal blocks of angular components from $\bW\mid R'>k$ with block size 4 by re-weighting empirical samples from $\bW \mid R'>1$ with important weights. 
From the pairwise spatio-temporal tail dependence coefficient plot shown in Figure S.3 in Supplementary Material A.1, we observe that the tail dependence coefficient stabilises for time lags beyond 3 or 4, suggesting that a block size of 4 adequately captures the extremal temporal dependence.
For each sampled block, we retain four temporally consecutive angular exceedances.
Using these temporal angular blocks in the extrapolation scheme, we estimate each target probability and then convert it to a frequency scale by multiplying by the number of observations in the corresponding climate model run.

We summarise the resulting point estimates along with bootstrap means, medians, and 95\% bootstrap confidence intervals in Table~\ref{tab:CTQall} across four model runs.
Comparing these with the empirical target values across 50 runs (CTQ1: 0.24, CTQ2: 0.20, CTQ3: 0.24), we observe that our estimates generally perform well for climate model runs 2, 3, and 4, whereas our fitted model shows a slight overestimation of tail probabilities in run 1. 
Furthermore, we see slightly more bias in CTQ3 estimates across all model runs compared to the 50-run empirical values, suggesting a full spatio-temporal extension could further improve the results.

\begin{table}[h!]
    \centering
    \begin{subtable}{0.9\linewidth}
    \centering
    \begin{tabular}{c|c|c|c|c}
         Run & Point estimate & Bootstrap - mean & Bootstrap - median & Bootstrap CIs\\ \hline
        1 & 0.523 & 0.530 &  0.530 & [0.382, 0.690]\\
        2 & 0.307 & 0.303 & 0.299 & [0.215, 0.410] \\
        3 & 0.215 & 0.230 & 0.225 & [0.153, 0.319] \\
        4 & 0.247 & 0.231 & 0.228 & [0.151, 0.328]
    \end{tabular}
    \caption{CTQ1}
    \label{tab:CTQ1_results}
    \end{subtable}

    \begin{subtable}{0.9\linewidth}
    \centering
    \begin{tabular}{c|c|c|c|c}
        Run & Point estimate & Bootstrap - mean & Bootstrap - median & Bootstrap CIs\\ \hline
        1 & 0.408 &  0.407 &  0.406 & [0.312, 0.505]\\
        2 & 0.138 & 0.122 & 0.121 & [0.088, 0.166]\\
        3 & 0.221 & 0.213&  0.210 & [0.157, 0.285] \\
        4 & 0.320 & 0.315 & 0.312 & [0.243, 0.393]
    \end{tabular}
    \caption{CTQ2}
    \label{tab:CTQ2_results}
    \end{subtable}

    \begin{subtable}{0.9\linewidth}
    \centering
    \begin{tabular}{c|c|c|c|c}
        Run & Point estimate & Bootstrap - mean & Bootstrap - median & Bootstrap CIs\\ \hline
        1 & 0.615 & 0.405 & 0.391 & [0.246, 0.619]\\
        2 & 0.297 & 0.139 & 0.132 & [0.060, 0.259]\\
        3 & 0.577 & 0.349 & 0.344 & [0.135, 0.622] \\
        4 & 0.355 & 0.357 & 0.343 & [0.200, 0.563]
    \end{tabular}
    \caption{CTQ3}
    \label{tab:CTQ3_results}
    \end{subtable}
    \caption{Estimated target quantities for CTQs 1-3 (top to bottom) across each climate model run.}
    \label{tab:CTQall}
\end{table}

We note a discrepancy between the results reported here and our submission for the data challenge rankings.
In that submission, the mean point estimates for CTQs across the four climate model runs, along with their 95\% bootstrap confidence intervals, were 5.06 (3.51, 5.78), 0.44 (0.23, 0.64), and 0.21 (0.08, 0.26), respectively.
The estimate for CTQ1 was overly high, raising concerns about the methodology.
To improve our approach, we implemented several updates in this work:
\begin{enumerate}
    \item Accounting for temporal and spatial non-stationarity:
    Our original semi-parametric marginal transformation did not account for temporal non-stationarity. We now preprocess the data to address temporal non-stationarity and obtain appropriate threshold values on the exponential scale.
    In addition, we replace the $L_1$-norm with the Euclidean distance for spatial lags and address anisotropy and spatial non-stationarity through spatial deformation.
    \item Selection of $k$:
    We adopt a practical procedure to select $k$, ensuring that a sufficient number of simulated points fall in the extreme sets of interest.
    \item Temporal block size:
    We adjust our initial temporal block size to 4 based on the observed behaviour of the spatio-temporal dependence coefficient, capturing the extremal temporal dependence appropriately. 
\end{enumerate}

\section{Discussion}
\label{sec:discussion}

Predicting extreme events is one of the central goals in statistics.
In this work, we address the challenge questions related to the spatial and/or temporal duration of extreme rainfall using the spatial geometric framework.
Our analysis focuses on a stationary spatial Gaussian-type process, which is a widely researched field in spatial statistics.
To estimate the temporal duration of extreme rainfall, we employ a bootstrap-based sampling procedure that retains the temporal information in the data.
This approach allows us to obtain reasonable probability estimates of how frequently extreme rainfall occurs simultaneously across multiple locations and the expected persistence of such events.

While our current analysis focuses on stationary spatial processes, the generalised Gaussian gauge function already provides flexibility to capture both space and temporal dependencies.
Incorporating a more flexible mixture of gauge functions could further improve the modelling of complex spatio-temporal extremes, which is particularly valuable when addressing real-world challenges in extreme event prediction.
A natural next step is to adopt a descriptive framework for spatio-temporal modelling suited to settings where there is limited knowledge of the underlying physical mechanisms driving the phenomena.

Extending the framework to fully spatio-temporal modelling is left for future work but could be achieved by adopting separable or non-separable correlation matrices in the generalised Gaussian gauge function.
This extension would allow the model to capture more realistic spatio-temporal effects.
Moreover, adapting the sampling approach to account for temporal dependence in the angular components would facilitate extrapolation to more extreme events over time.
While our focus here has been on the conditional distribution of the radial component, modelling of the angular distribution in a spatio-temporal context could additionally enrich the quality of the model, especially if higher dimensional settings are to be considered; such an extension also represents an interesting direction for future research.

\section*{Declarations}

\subsection*{Acknowledgements}
This paper is based on work completed while KG was part of the EPSRC funded STOR-i Centre for Doctoral Training EP/S022252/1.
LK was part of EPSRC funded project EP/W524438/1, RC was supported by EPSRC grant EP/W523811/1, and JL was supported by EPSRC grant EP/X010449/1.

\subsection*{Code Availability}
Code supporting the findings of this study is available from the corresponding author upon reasonable request.

\subsection*{Data Availability}
The data for the EVA (2025) Conference Data Challenge has been made publicly available at \url{https://eva2025.unc.edu/data-challenge/} by the organising committee. 

\subsection*{Competing Interests}
The authors have no relevant financial or non-financial interests to disclose.

\subsection*{Authors' Contributions}
All authors contributed equally to this work.

\subsection*{Supplementary Information}
This manuscript is accompanied by the document \textbf{Supplementary Material for ``Analysing Extreme Rainfall via a Geometric Framework''}.

\bibliography{bibliography}

\end{document}